# Critical role of next-nearest-neighbor interlayer interaction in magnetic behavior of magnetic/nonmagnetic multilayers

Sunjae Chung[1], Sangyep Lee[1], Taehee Yoo[1], Hakjoon Lee[1], J.-H. Chung[1], M. S. Choi[1], Sanghoon Lee[1]*, X. Liu[2], J. K. Furdyna[2], Jae-Ho Han[3], Hyun-Woo Lee[3], and Kyung-Jin Lee[4]

[1] Department of Physics, Korea University, Seoul, 136-701, KOREA
[2] Department of Physics, University of Notre Dame, Notre Dame, Indiana 46556, USA
[3] PCTP and Department of Physics, Pohang University of Science and Technology, Pohang, Kyungbuk 790-784, KOREA
[4] Department of Materials Science and Engineering, Korea University, Seoul 136-701, KOREA

* Email : slee3@korea.ac.kr

We report magnetoresistance data in magnetic semiconductor multilayers, which exhibit a clear step-wise behavior as a function of external field. We attribute this highly non-trivial step-wise behavior to next-nearest-neighbor interlayer exchange coupling. Our microscopic calculation suggests that this next-nearest-neighbor coupling can be as large as 24% of the nearest-neighbor coupling. It is argued that such unusually long-range interaction is made possible by the quasi-one-dimensional nature of the system and by the long Fermi wavelength characteristic of magnetic semiconductors.

73.50.-h, 75.50.Pp, 75.47.De

The interlayer exchange interaction [1] in magnetic multilayers gives rise to giant magnetoresistance, (GMR) [2, 3] and can thus lead to new breakthroughs in spin-electronics.[4] It is therefore important to understand the fundamental principles behind such exchange interaction and to investigate its physical behaviors under different circumstances. Indeed, many studies have already been carried out on the interlayer exchange interaction in multilayer systems with various ferromagnetic materials [5-9], ranging from the dependence of the interlayer coupling strength on structural parameters to the oscillation between ferromagnetic and antiferromagnetic coupling [1, 10, 11]. However, all these previous studies focused only on nearest-neighbor (NN) interactions between the magnetic layers.

In this Letter we report the observation of a striking step-wise behavior in magnetoresistance, which cannot be explained in the frame of any existing theories based only on nearest-neighbor coupling, thus providing clear evidence that next-nearest-neighbor (NNN) interactions are significant. Based on microscopic calculations, we show below that the NNN interaction is of critical importance in magnetic semiconductor multilayers, where the Fermi wavelength (~ 4 nm in our GaMnAs samples) is significantly longer than in metallic multilayers, becoming comparable to the width of the layers [12-18].

We note that the effect of next-nearest-neighbor coupling is most pronounced when the interlayer exchange coupling (IEC) is antiferromagnetic (AFM).[19-22] It turns out that the AFM IEC in GaMnAs based systems can be achieved in only a very narrow parameter window of carrier concentrations and structural dimensions. [16, 23-25] In this study we therefore focus on magnetization reversal observed in a [GaMnAs/GaAs:Be]$_{10}$ multilayer with GaMnAs and GaAs:Be thicknesses of 7 nm and 3.5 nm, respectively, in which AFM IEC is very clearly realized. We note, however, that we have observed similar AFM IEC behavior in other multilayer structures as well (see Supplementary Material). The detailed growth process of such GaMnAs/GaAs multilayers and MR measurement are described elsewhere. [16]

Figure 1a shows magnetoresistance (MR) data obtained at 35 K during magnetization reversal with the applied magnetic field oriented near the [110] direction. The curved arrows indicate the directions of field sweep, and the solid (blue) and open (red) circles represent the data obtained during the down-scan (i.e., field sweep from positive saturation to negative saturation) and the up-scan (field sweep from negative saturation to positive saturation) of the field, respectively. In discussing these and other data, we will refer to the process of increasing the field in either direction (which eventually aligns the magnetization of all GaMnAs layers parallel) as the "saturation" process or sweep; and we will refer to the process of recovering the initial antiparallel (AFM) configuration by reducing the field to zero as the "restoring" process. In addition to the basic GMR-like effect, we now see many clearly resolved transition steps in both the saturation and the restoring sweeps. One could at first glance suspect that the multiple transition steps which we observe are caused by unintended variations of magnetic properties (specifically, magnetic anisotropy) between individual GaMnAs layers. However, the observed characteristics, such as the regular spacing of the transition steps, the large difference of the transition fields (about 40 Oe between the first and the last transitions), and the different sequence of transition steps in the saturation (five-steps) and restoring sweeps (four-steps) cannot be explained by an accidental variation of GaMnAs layers (which are grown consecutively in a single growth run under the same growth conditions). We can therefore rule out the possibility that the observed behavior is caused by unintended differences between the GaMnAs layers comprising the multilayer.

It is remarkable that the number of transition steps (five) which appear in the saturation process exactly matches the number of GaMnAs layers whose magnetization is initially aligned opposite to the direction of the applied field in the AFM alignment of the multilayer at zero field. This implies that the magnetizations of those GaMnAs layers reverses one by one as the field increases toward saturation, as shown in Fig. 1a. When the magnetic multilayer is cooled below $T_C$ in the absence of an external field, a system consisting of 10 AFM-coupled GaMnAs layers can be thought of as forming two types of anti-parallel spin configurations, AFM1 ($\downarrow\uparrow\cdots\downarrow\uparrow$) or AFM2 ($\uparrow\downarrow\cdots\uparrow\downarrow$), as schematically shown in Fig. 1 for $H = 0$, in which arrows indicate the directions of magnetization, and the progression of the arrows from left to right corresponds to the order of the GaMnAs layers from the bottom to the top of the multilayer.

Although it is obvious that the first transition in the saturation process corresponds to the flip of magnetization in the outermost GaMnAs layer, it is difficult to identify the sequence of transitions of the inner GaMnAs layers, since the order of such flips cannot be determined by considering NN IEC alone, but must involve additional interaction terms that differentiate the strengths of IEC acting on specific interior GaMnAs layers. The most likely interaction for removing the energy degeneracy of magnetization alignments of the interior GaMnAs layers determined by NN AFM IEC is the next-nearest-neighbor (NNN) IEC. As already mentioned, this process is likely to come into play in the present case, since the Fermi wavelength of our GaMnAs samples is about 4 nm long. [14, 16]

We therefore performed a microscopic calculation to obtain the relative strength of NNN IEC with respect to that of NN IEC by evaluating the magnitude of spin torque between NNN the magnetic layers of the multilayer (See Supplementary Material). The results show that the strength of NNN IEC can be as large as 24% of NN IEC strength. This surprisingly large value of NNN IEC may be attributed to (i) the quasi-one-dimensionality and (ii) the small thickness of the layers, as follows. (i) In bulk solids, the Ruderman–Kittel–Kasuya–Yosida (RKKY) type exchange coupling is known to decrease (at large scales) as $1/r^3$, where $r$ is the distance between the interacting spins. However, for a quasi-one-dimensional system, such as our multilayer sample it was shown [26-29] that the IEC strength between magnetic layers decreases only as $1/r^2$, i.e., much slower. (ii) The thickness of each layer in our [GaMnAs/GaAs:Be]$_{10}$ multilayer system is comparable to the Fermi wavelength, and the exchange coupling does not follow the simple power-law decay.

In discussing the magnetization reversal behavior of the GaMnAs/GaAs multilayer we focus on understanding of the multiple transition step behavior – the key feature of this experiment – rather than on quantitative details of the entire hysteresis, that would of necessity include domain wall pinning and magnetic anisotropy of each GaMnAs layers.

We, therefore, consider only IEC energy of the multilayer and formulate the IEC energy $E$ of the multilayer in the presence of a magnetic field $H$ by including NNN IEC contributions. For the case of a 10-period multilayer such as our [GaMnAs/GaAs:Be]$_{10}$ SL the coupling energy can then be expressed in the form

$$E = \sum_{i=1}^{9} J_1 (M_i \cdot M_{i+1}) + \sum_{i=1}^{8} J_2 (M_i \cdot M_{i+2}) - \sum_{i=1}^{10} H \cdot M_i \qquad (1),$$

where $J_1$ and $J_2$ are the NN and NNN IEC constants, respectively, and $M_i$ is the magnetization of the $i_{th}$ GaMnAs layer.

To bring out the role of NNN IEC, it is useful to first calculate the magnetic field dependence of the IEC energy for various spin configurations *without* the NNN IEC contribution (i.e., without the 2$^{nd}$ term in Eq. (1)). The results are shown in Fig. 2a, and clearly illustrate the linear dependences of the coupling energies on the field for each spin configuration. The figure shows that these energy dependences have only two crossing points, corresponding to two critical fields at which these energies have multiple degeneracies. Specifically, the first (lower) critical field corresponds to the magnetization flip of the *outermost* magnetic layer that was initially aligned opposite to the applied field; and the second crossing point occurs when the magnetizations of *all interior magnetic layers* initially aligned opposite to the field flip their orientations. Importantly, in the NN IEC scenario this happens at a single field for all internal layers. Thus, in this scenario only two transitions occur in both the saturation and the restoring processes, as has indeed been observed in metallic magnetic multilayers. [6]

We will now include the NNN IEC in the calculation. Although we will assume the strength of NNN IEC to be J2 = 0.25J1 in our calculation, qualitatively similar results are obtained with other values of J2 if the ratio J2/J1 is not negligible small. The most remarkable effect of including the NNN IEC is that it removes the energy degeneracy of the second crossing point seen in the NN IEC picture (i.e., that seen in Fig. 2a), as will be discussed in detail in connection with Fig. 2b.

Calculation of the lowest energy state according to Eq. (1) shows that, as the field is swept, many transition paths with stable spin configurations may occur, each of these paths leading to magnetization reversal with multiple intermediate states. The results of this calculation alone are not sufficient to unambiguously determine the actual spin configurations that occur in the reversal process. Fortunately, however, the experiment provides several additional conditions that narrow down the exact spin configuration corresponding to each resistance state (i.e., each step) observed during magnetization reversal. For example, it is remarkable that in going from negative to positive saturation (up-scan) and vice versa (down-scan) we see exactly the same 4-step restoration and 5-step saturation paths, as shown in Fig. 1a. This perfectly symmetric behavior of MR seen in the up- and down-scans of the field indicates that the AFM spin configuration at zero field switches from AFM1 and AFM2 (shown on top of Fig. 1a) on completing the 5-step path to saturation and the 4-step return (i.e., AFM1 → FM1 → AFM2; and then AFM2 → FM2 → AFM1). This experimental observation then to establishes the following two constraints.

*1. The system always experiences complete antiparallel spin configuration (i.e., either AFM1 ($\downarrow\uparrow\cdots\downarrow\uparrow$) or AFM2 ($\uparrow\downarrow\cdots\uparrow\downarrow$)) at zero field as the field is scanned from saturation in one direction to saturation in the opposite direction.*

*2. The antiparallel spin configuration at zero field always switches between AFM1 ($\downarrow\uparrow\cdots\downarrow\uparrow$) and AFM2 ($\uparrow\downarrow\cdots\uparrow\downarrow$) when the system is restored after experiencing a 5-step saturation process.*

Furthermore, the fact that our MR measurements are performed in the current in-plane configuration automatically identifies the location of the first transition, i.e., whether it occurs in the bottom or the top GaMnAs layer of the SL, since the current flow in this configuration will be systematically less sensitive to magnetization changes in GaMnAs layers located away from the surface of the SL (i.e., from the contacts to the sample). This leads to the third constraint.

*3. The change of resistance (measured by the voltage drop) will be the smallest for the flip of magnetization in the GaMnAs layer at the bottom of the structure*

Satisfying these three conditions uniquely determines the sequence of spin configurations (from among many possible sequences) in the course of magnetization reversal. The magnetic field dependences of energies for the resulting spin configurations for the down-scan case are shown in Fig. 2b. The most interesting feature of Fig. 2b is the splitting of the second crossing point in Fig. 2a (i.e., that obtained by considering only the NN IEC) into many different crossing points. Now 4 and 5 crossing points appear in the restoring and the saturation process, respectively, (as indeed seen in the data in Fig. 1a) each crossing point corresponding to flip of magnetization in a particular GaMnAs layer.

Since the value of MR in the magnetic multilayer is a reflection of electron scattering between neighboring anti-parallel magnetic layers, we count the number of anti-parallel neighboring layer pairs occurring in the different spin configurations as the magnetization is being reversed, and plot the calculated field dependence of this number in Fig. 1b for comparison with the MR experiment. As in the case of Fig. 2b, for clarity, we only plot the results for the down-scan, i.e., FM1→AFM2→FM2. The magnetization alignments within the structure that lead to each resistance state during the field scans are also schematically shown in Fig. 1b. As one can see, the magnetization reversal path obtained by calculating the lowest coupling energy with the three experimental constraints taken into account reproduces the observed 4-step restoration and the 5-step saturation processes. This implies that the obtained sequence of spin configurations corresponds to the sequence of resistance states observed in the field sweeps. Since this sequence of spin configurations during magnetization reversal is obtained by considering NNN IEC, these results serve to underscore that the NNN interaction is the key factor for understanding the details of magnetization reversal in FM semiconductor multilayers such as our [GaMnAs/GaAs:Be]$_{10}$ SL.

It is interesting to note that the restoring process (from saturation to full AFM order) undergoes a 4-step transition sequence, in contrast to the 5 steps which we observe when progress from full AFM order to full FM saturation. Although this is at first glance quite surprising, it can be also understood from the results of calculation shown in the inset in Fig.

1b, where the field dependence of magnetic energies is plotted for the three spin configuration states (i.e., ↑↑↑↑↑↑↑↑↑, ↑↑↑↑↑↑↓↑↑↑, and ↑↑↓↑↑↑↓↑↑↑, designated in the figure as FM1, R4, and R3, respectively). A completely parallel configuration has the lowest energy in the region of magnetic fields larger than the field at which the FM1, R4, and R3 states have same energy, which we designate as $H_c$.

Surprisingly, as the field is reduced below $H_c$, the magnetization configuration with two anti-parallel layers, R3, acquires a lower energy than the configuration with one anti-parallel layer, R4, even though R3 is magnetically further from FM1. Thus the system makes a transition at $H_c$ *directly* from a completely parallel configuration FM1 to the configuration with *two* anti-parallel layers, R3. That is, the state with only one antiparallel layer simply does not occur during restoration, so that this simultaneous flip of magnetization in *two* GaMnAs layers at the beginning of the restoring process thus results in the 4-step restoring sequence, as observed in experiment. For additional details, see Supplementary Materials.

Finally, the explanation of the various transition patterns (such as the number of steps during a given reversal process and the spin configuration of each state) is made possible in terms of the contributions of NNN IEC, which was estimated to have an approximate strength of $J_2 = 0.25 J_1$. This contribution of NNN IEC is quite pronounced in GaMnAs/GaAs:Be multilayers because of the long range character of IEC in this semiconductor-based system, thus providing a powerful tool for investigating higher order IEC effects by experiment. We hope that this direct detection of NNN IEC in our multilayer will stimulate investigation of higher order IEC effects in other magnetic multilayer systems. Indeed, a study on the Co/Cu/Co/Cu$_{cap}$ system has already shown that the IEC between two Co layers depends on the thickness of the Cu capping layer. [30] Such non-magnetic capping-layer dependence of IEC suggests the possibility of NNN IEC in multilayers consisting of more than 3 magnetic/non-magnetic pairs. Furthermore, the presence of NNN IEC in a magnetic multilayer structure also opens a new opportunity for realizing multiple stable states that can be used for multinary information storage and/or logic devices, in which spin configurations established by IEC are controlled via electrical methods, such as injecting charge carriers into the non-magnetic spacers of the multilayer structure.

This research was supported by the Converging Research Center Program through the Ministry of Education, Science and Technology (2011K000786); by a National Research Foundation of Korea (NRF) grant, funded by the Korean government (MEST) (Nos. 2010-0025880); and by the National Science Foundation Grants DMR10-05851. JHH and HWL acknowledge the financial support by the NRF (Nos. 2010-0014109, 2011-0030784).

**References**

[1] P. Grünberg, R. Schreiber, Y. Pang, M. B. Brodsky, and H. Sowers, Phys. Rev. Lett. **57**, 2442 (1986).
[2] M. N. Baibich, J. M. Broto, A. Fert, F. N. Van Dau, F. Petroff, P. Etienne, G. Creuzet, A. Friederich, and J. Chazelas, Phys. Rev. Lett. **61**, 2472 (1988).
[3] G. Binasch, P. Grünberg, F. Saurenbach, and W. Zinn, Phys. Rev. B **39**, 4828 (1989).
[4] S. A. Wolf, D. D. Awschalom, R. A. Buhrman, J. M. Daughton, S. von Molnar, M. L. Roukes, A. Y. Chtchelkanova, and D. M. Treger, Science **294**, 1488 (2001).
[5] J. Unguris, R. J. Celotta, and D. T. Pierce, Phys. Rev. Lett. **67**, 140 (1991).
[6] P. J. H. Bloemen, H. W. van Kesteren, H. J. M. Swagten, and W. J. M. de Jonge, Phys. Rev. B **50**, 13505 (1994).
[7] J. A. Borchers, J. A. Dura, J. Unguris, D. Tulchinsky, M. H. Kelley, C. F. Majkrzak, S. Y. Hsu, R. Loloee, W. P. Pratt, and J. Bass, Phys. Rev. Lett. **82**, 2796 (1999).
[8] H. Kepa, V. K. Le, C. M. Brown, M. Sawicki, J. K. Furdyna, T. M. Giebultowicz, and T. Dietl, Phys. Rev. Lett. **91**, 087205 (2003).
[9] J. J. Rhyne, J. Lin, J. K. Furdyna, and T. M. Giebultowicz, Journal of Magnetism and Magnetic Materials **177-181**, 1195 (1998).
[10] S. S. P. Parkin, Phys. Rev. Lett. **67**, 3598 (1991).
[11] S. S. P. Parkin, N. More, and K. P. Roche, Phys. Rev. Lett. **64**, 2304 (1990).
[12] J. H. Chung, S. J. Chung, S. Lee, B. J. Kirby, J. A. Borchers, Y. J. Cho, X. Liu, and J. K. Furdyna, Phys. Rev. Lett. **101**, 237202 (2008).
[13] J. Chung, Y. Song, T. Yoo, S. J. Chung, S. Lee, B. J. Kirby, X. Liu, and J. K. Furdyna, J. Appl. Phys. **110**, 013912 (2011).
[14] J. Leiner, H. Lee, T. Yoo, S. Lee, B. J. Kirby, K. Tivakornsasithorn, X. Liu, J. K. Furdyna, and M. Dobrowolska, Phys. Rev. B **82**, 195205 (2010).
[15] J. Leiner, K. Tivakornsasithorn, X. Liu, J. K. Furdyna, M. Dobrowolska, B. J. Kirby, H. Lee, T. Yoo, and S. Lee, J. Appl. Phys. **109**, 07C307 (2011).
[16] S. Chung, S. Lee, J. H. Chung, T. Yoo, H. Lee, B. Kirby, X. Liu, and J. K. Furdyna, Phys. Rev. B **82**, 054420 (2010).
[17] T. Dietl, H. Ohno, F. Matsukura, J. Cibert, and D. Ferrand, Science **287**, 1019 (2000).
[18] T. Dietl, H. Ohno, and F. Matsukura, Phys. Rev. B **63**, 195205 (2001).
[19] D. Chiba, N. Akiba, F. Matsukura, Y. Ohno, and H. Ohno, Appl. Phys. Lett. **77**, 1873 (2000).
[20] B. J. Kirby, J. A. Borchers, X. Liu, Z. Ge, Y. J. Cho, M. Dobrowolska, and J. K. Furdyna, Phys. Rev. B **76**, 205316 (2007).
[21] H. Kępa, J. Kutner-Pielaszek, A. Twardowski, C. F. Majkrzak, J. Sadowski, T. Story, and T. M. Giebultowicz, Phys. Rev. B **64**, 121302 (2001).
[22] N. Akiba, F. Matsukura, A. Shen, Y. Ohno, H. Ohno, A. Oiwa, S. Katsumoto, and Y. Iye, Appl. Phys. Lett. **73**, 2122 (1998).
[23] P. Sankowski, and P. Kacman, Phys. Rev. B **71**, 201303 (2005).
[24] A. D. Giddings, T. Jungwirth, and B. L. Gallagher, Phys. Rev. B **78**, 165312 (2008).
[25] K. Szalowski, and T. Balcerzak, Phys. Rev. B **79**, 214430 (2009).
[26] C. Kittel, edited by F. Seitz, D. Turnbull, and H. Ehrenreich (academic Press, 1969), pp. 1.
[27] Y. Yafet, Phys. Rev. B **36**, 3948 (1987).
[28] P. Bruno, and C. Chappert, Phys. Rev. Lett. **67**, 1602 (1991).
[29] P. Bruno, and C. Chappert, Phys. Rev. B **46**, 261 (1992).
[30] J. J. d. Vries, A. A. P. Schudelaro, R. Jungblut, P. J. H. Bloemen, A. Reinders, J.

Kohlhepp, R. Coehoorn, and W. J. M. de Jonge, Phys. Rev. Lett. **75**, 4306 (1995).
**Figures**

Fig. 1 The process of magnetization reversal in the [GaMnAs/GaAs:Be]$_{10}$ multilayer system. (a) Magnetoresistance is measured at 35 K as the field is cycled between -100 Oe and 100 Oe. The down- and up-scans, shown by solid (blue) and open (red) circles, respectively, have a completely symmetric behavior. Two types of fully antiferromagnetic spin configurations between the GaMnAs layers, AFM1 and AFM2, can be realized at zero field, as shown schematically by the vertical arrows. Each field scan (down or up) contains a 4-step restoring and a 5-step saturation processes, with resistance plateaus marked as R1-R4 and S1-S4. (b) The number of pairs with AFM alignment between adjacent GaMnAs layers in the multilayer is obtained by minimizing the IEC energy given by Eq. (1) during the down-scan of the field. The field is scaled in terms of the NN IEC strength $J_1$. The reversal process determined from calculation using Eq. (1) clearly shows a 4-step restoring and a 5-step saturation process, similar to that observed in the MR experiment shown in the upper panel. The crossing of the calculated energies for the R3, R4, and FM1 states is shown in the inset. The spin configuration corresponding to each plateau in the field scan is indicated schematically by vertical arrows.

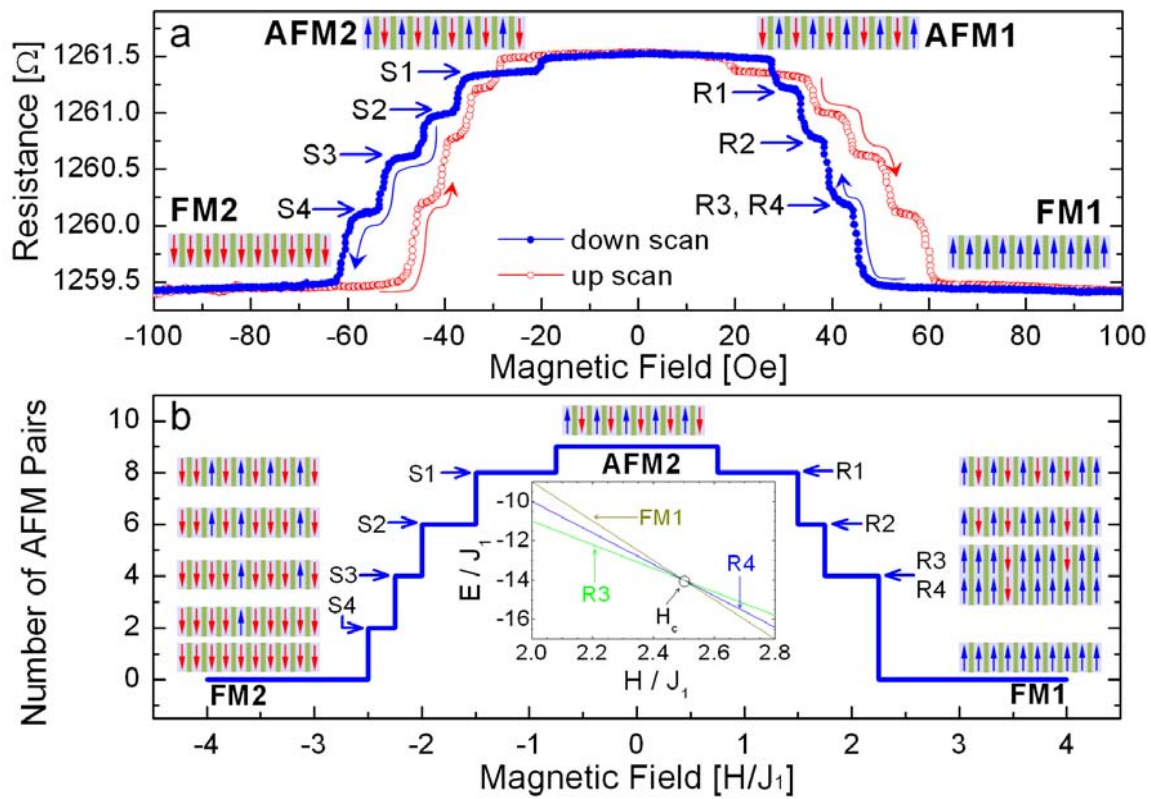

Fig. 2 Calculated magnetic field dependence of energies for different spin configurations in a 10-period multilayer. (a) Upper panel shows the calculation with only NN IEC included. As indicated by the circles, only two crossing points are present for each field sweep direction. Note that in this case many spin configurations involving inner GaMnAs layers are degenerate in energy, and that the spin configurations indicated by the vertical arrows show only one representative configuration. (b) The lower panel shows energies calculated with both NN and NNN IEC included. The specific spin configurations are determined from the three experimental constraints, as explained in the text. The results are shown only for the down-sweep of the field (i.e., from positive-field saturation to negative-field saturation). The results clearly show 4 and 5 crossing points, corresponding to the 4-step restoring and the 5-step saturation processes, respectively, as seen in the MR experiments. The dotted arrow indicates the sweeping direction of the field (down-sweep).

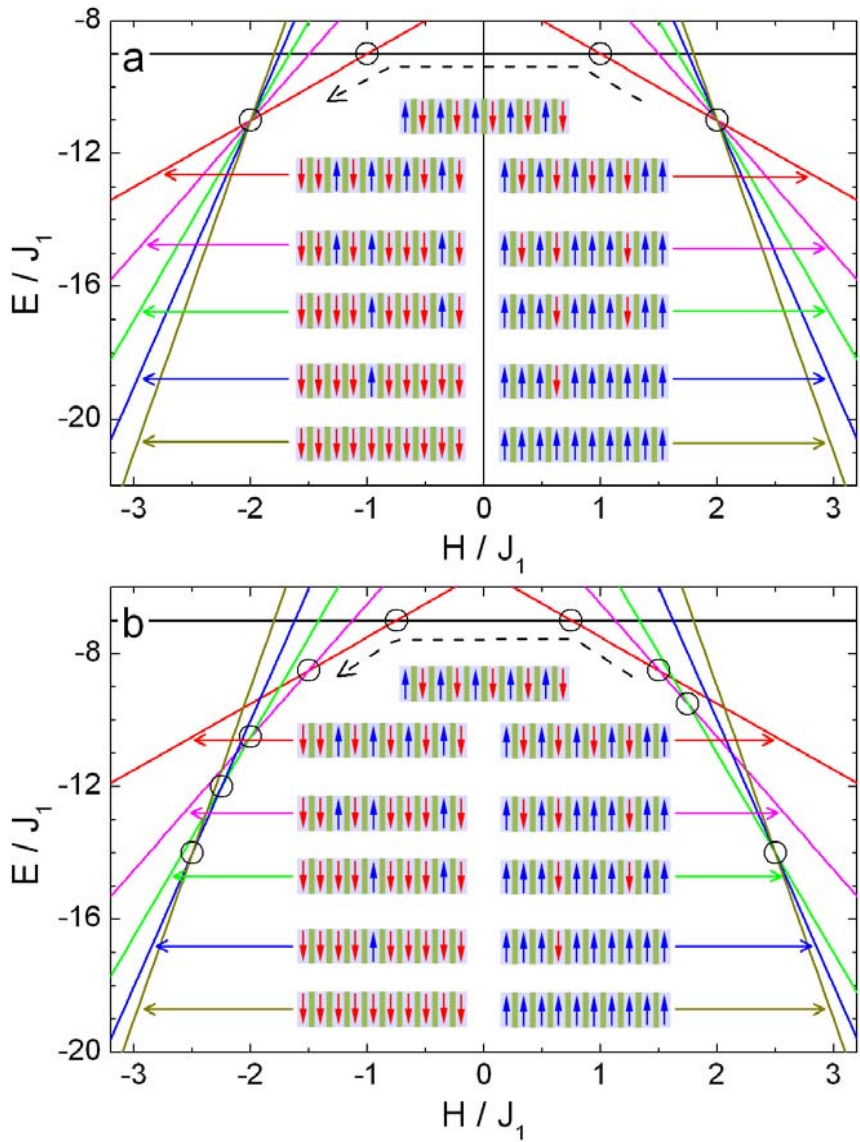

# Supplementary Materials for

# Critical role of next-nearest-neighbor interlayer interaction in magnetic behavior of (Ga,Mn)As multilayers

Sunjae Chung, Sangyep Lee, Taehee Yoo, Hakjoon Lee, J.-H. Chung, M. S. Choi, Sanghoon Lee[*], X. Liu, J. K. Furdyna, Jae-Ho Han, Hyun-Woo Lee, and Kyung-Jin Lee

Correspondence to:   slee3@korea.ac.kr

- **Magnetoreistance of the of GaMnAs/GaAs:Be multilayers**

In addition to the multilayer on which we focus in the main text, we also investigated other multilayer specimens with different structural parameters. Here we present results for three samples listed in Table S1. The magnetoresistance data taken at 30 K on these specimens are plotted in Fig. S1. Sample 1 shows that the value of resistance is not a maximum at zero field, an abrupt increase of resistance occurring when the applied field is reversed. This behavior is typical for anisotropic magnetoresistance (AMR), and is also observed on single layers of GaMnAs grown on GaAs substrates. [1, 2] This indicates that the interaction between the GaMnAs layers in Sample 1 is ferromagnetic (FM IEC). In contrast, Samples 2 and 3 clearly show a maximum resistance at zero field during the field scans. Such recovery of the resistance maximum at zero field during field cycling is typically observed in magnetic multilayer systems with AFM IEC, and represents the mechanism responsible for giant magnetoresistance (GMR). [3] The observation of this GMR-like effect in a series of GaMnAs-based multilayers with different structural parameters provides additional evidence that the AFM IEC such as that discussed in the main text is a general characteristic, and can be reproducibly realized by appropriate selection of structural parameters.

- **Microscopic calculation of inter layer exchange coupling**

The aim of this supplementary section is to estimate the magnitude of the NNN IEC constant $J_2$ and to compare it to the NN IEC constant $J_1$. Theoretical calculation will be carried out for a model multilayer FNFNF shown in Fig. S2, where F and N denotes ferromagnetic and nonmagnetic layers, respectively. For this model system, the NNN IEC implies that the spin information can be exchanged between ferromagnetic layers F1 and F3 which are, respectively, on the left and on the right of layer F2. Before presenting our explicit evaluation of the ratio $J_2/J_1$, we will first discuss the factors affecting this ratio. To carry spin information exchange between F1 and F3, charge carriers (holes in our experimental situation) should travel through not only the left and the right nonmagnetic layers (N1 and N2), but also through the intervening ferromagnetic layer F2. Within N1 and N2, the spin information will oscillate and decay according to the RKKY formula, which predicts the decay of IEC to be inversely proportional to the square of the separation between the ferromagnetic layers. [4] Within ferromagnetic layers, on the other hand, spin information propagation can be more complicated since the ferromagnetism itself may generate loss of spin information of the itinerant carriers.

To estimate the actual $J_2/J_1$ ratio, one needs to assess the spin information loss within F2, which will depend sensitively on the angle between the spin of the carrier and the direction of magnetization in the ferromagnetic layer. The transverse spin component (i.e., the component perpendicular to the magnetization direction in the FM layer) of the carrier is strongly suppressed in F2 due to a number of mechanisms, [5] including (i) spin dephasing within ferromagnets, (ii) spin-dependent reflection and transmission at the interface between nonmagnets and ferromagnets, and (iii) rotation of the carrier spin direction upon reflection and transmission. Figure S3 illustrates the spin dephasing mechanism in the ferromagnet. In case of metallic multilayer systems consisting of nonmagnetic metals and metallic ferromagnets, the combined effect of these mechanisms [5] results in almost complete suppression of the transverse spin component after carriers enter the FM layer even by a few lattice constants. We expect that similar phenomena occur in semiconducting multilayer systems and thus the transverse spin component cannot contribute to the NNN IEC.

On the other hand, the longitudinal spin component, i.e., the component parallel or antiparallel to the magnetization direction of the layer, can be maintained over a much longer distance. All those mechanisms[5] that suppress the transverse component do not eliminate the longitudinal spin component, and thus the longitudinal spin component may be maintained over longer distances.

This reasoning leads to the following form of the NNN IEC:

$$\text{NNN IEC} = +\frac{J_2}{M_s^2}(\mathbf{M}_1 \cdot \mathbf{M}_2)(\mathbf{M}_2 \cdot \mathbf{M}_3), \tag{S1}$$

where $\mathbf{M}_1$, $\mathbf{M}_2$, $\mathbf{M}_3$ represent the magnetizations of layers F1, F2, and F3, respectively, in Fig. S2 and $M_s=|\mathbf{M}_1|=|\mathbf{M}_2|=|\mathbf{M}_3|$ is the value of saturation magnetization in each of the layers. This form implies that only those components of $\mathbf{M}_1$ and $\mathbf{M}_3$ which are parallel (or antiparallel) to $\mathbf{M}_2$ contribute to the NNN IEC between F1 and F3. In the special case where the system has a uniaxial magnetic anisotropy, and $\mathbf{M}_1$, $\mathbf{M}_2$ and $\mathbf{M}_3$ are either parallel or anti-parallel to each other, Eq. (S1) reduces to

$$\text{NNN IEC} = +J_2(\mathbf{M}_1 \cdot \mathbf{M}_3), \tag{S2}$$

where $\mathbf{M}_2$ does not appear, since $(\ldots \cdot \mathbf{M}_2)(\mathbf{M}_2 \cdot \ldots)/M_s^2=1$. Equation (S2) has the form that appears in the NNN IEC term of Eq. (1) in the main text.

To verify Eq. (S1) by microscopic calculation, we take advantage of the relation that the variation of the total IEC energy with respect to the directional variation of $\mathbf{M}_i$ is related to the spin torque [6] acting on $F_i$ ($i$=1,2,3). To be specific, the spin torque for a given magnetic configuration is a function of the current flowing perpendicular to the layers, and spin torque in the zero current limit [7] is determined by the variation of the total IEC energy near the given magnetic configuration. To illustrate this relation, suppose that the total IEC energy $E$ is as follows,

$$E = J_1(\mathbf{M}_1 \cdot \mathbf{M}_2 + \mathbf{M}_2 \cdot \mathbf{M}_3) + \frac{J_2}{M_s^2}(\mathbf{M}_1 \cdot \mathbf{M}_2)(\mathbf{M}_2 \cdot \mathbf{M}_3), \tag{S3}$$

where the first term represents the NN IEC. For simplicity, suppose also that $\mathbf{M}_1$, $\mathbf{M}_2$, $\mathbf{M}_3$ vectors lie within the same two-dimensional (2D) plane. Then the change of $E$ caused by the rotation of $\mathbf{M}_3$ within the 2D plane amounts to the spin torque acting on F3 with the torque direction perpendicular to the 2D plane, and with the magnitude of the torque $\tau_3$ given by

$$\begin{aligned}\tau_3 &= -\frac{\partial E}{\partial \theta_{23}} = J_1 M_s^2 \sin\theta_{23} + \frac{J_2}{M_s^2}(\mathbf{M}_1 \cdot \mathbf{M}_2) M_s^2 \sin\theta_{23} \\ &= \left(J_1 + J_2 \frac{\mathbf{M}_1 \cdot \mathbf{M}_2}{M_s^2}\right) M_s^2 \sin\theta_{23}\end{aligned}, \tag{S4}$$

where $\theta_{23}$ is the angle between $\mathbf{M}_2$ and $\mathbf{M}_3$. Thus the calculation of $\tau_3$ for various values of

**M**$_1$· **M**$_2$ allows separate evaluation of $J_1$ and $J_2$.

The calculation then proceeds as follows. We consider the FNFNF system shown in Fig. **S2**, where the magnetization directions **m**$_i$=|**M**$_i$|/$M_s$ ($i$=1,2,3) in F1, F2, and F3 are denoted by the arrows, $d_1$ and $d_2$ are the thicknesses of N1 and N2, and $L$ is the thickness of F2. F1 and F3 are for simplicity assumed to be semi-infinite. Also the layers are assumed to be infinite in the $x$- and $z$-directions. To describe the carrier dynamics in this system, we adopt an independent carrier model (i.e., there is no interaction between the carriers) and solve the following Schrödinger equation for the carrier wave function $\Psi=(\Psi_\uparrow,\Psi_\downarrow)^T$,

$$\left(\frac{\mathbf{p}^2}{2m}+U\right)\begin{pmatrix}\Psi_\uparrow\\\Psi_\downarrow\end{pmatrix}=E\begin{pmatrix}\Psi_\uparrow\\\Psi_\downarrow\end{pmatrix}, \tag{S5}$$

where $m$ is the effective mass of the carriers and $U$ is a spin-dependent potential representing the exchange interaction between the carrier spin and the magnetization. The effective mass approximation is well suited for semiconductor systems, since the energies of the carriers are near the band edge. For the FNFNF structure shown in Fig. S2, $U$ is given by [8]

$$U=\begin{cases}\frac{\Delta}{2}\sigma\cdot\mathbf{m}_1 & \text{within F1}\\ \frac{\Delta}{2}\sigma\cdot\mathbf{m}_2 & \text{within F2}\\ \frac{\Delta}{2}\sigma\cdot\mathbf{m}_3 & \text{within F3}\\ 0 & \text{within N1, N2}\end{cases}, \tag{S6}$$

where $\Delta$ represents the strength of the exchange interaction between carrier spins and the magnetizations **m**$_i$ of individual layers. The energy splitting between the majority and minority spins is thus given by $\Delta$. Note that the spin dephasing is explicitly taken into account in the Schrödinger equation through $U$. On the other hand, impurity potentials which could generate spin flip scattering are not included in the Schrödinger equation, based on the assumption that the longitudinal spin relaxation length sufficiently exceeds $d_1+L+d_2$.

The evaluation of the spin torque is now straightforward. By using the eigenstate wave function of the Schrödinger equation, one first calculates the total spin current density $\mathbf{J}^\alpha_{\text{spin}}(\mathbf{r})$ of the spin component $\alpha$ ($\alpha$=$x, y, z$),

$$\mathbf{J}^\alpha_{\text{spin}}(\mathbf{r})=\sum_{k\ (E_k<E_F)}\sum_{s,s'=\uparrow,\downarrow}\text{Re}\left[\{\Psi^*_k(\mathbf{r})\}_s\left\{\frac{\hbar}{2}\sigma^\alpha_{s,s'}\otimes\frac{\mathbf{p}}{m}\right\}\{\Psi_k(\mathbf{r})\}_{s'}\right], \tag{S7}$$

where $\Psi_k$ represents the $k$-th eigenstate with energy $E_k$; $\{\Psi_k\}_\uparrow$ and $\{\Psi_k\}_\downarrow$ are the spin-up and spin-down components of $\Psi_k$; and the summation over $k$ is restricted to the states below the Fermi energy $E_F$. With this, one can now calculate the spin torque **T**$_3$ acting on layer F3 by using the expression [9]

$$\mathbf{T}_3\cdot\hat{\mathbf{x}}_\alpha=-\int_{S_3}d\mathbf{S}_3\cdot\mathbf{J}^\alpha_{\text{spin}}, \tag{S8}$$

where $S_3$ is the surface enclosing F3, $d\mathbf{S}_3$ is the surface element of $S_3$ with the direction of the normal outgoing relative to the surface, and $\hat{\mathbf{x}}_\alpha$ is the unit vector along the $\alpha$ direction. The spin torque calculation based on carrier transport across the multilayers is the same for both the valence [8] or the impurity band [10] models, and involving only slightly different band parameters. Thus the conclusions described here are applicable to both models.

For calculating **T**$_3$ we use the following parameters taken from the valence band model: $d_1=d_2=3.5$ nm, $L=7.0$ nm, $E_F= 155$ meV, $\Delta= 140$ meV, and $m= 0.5m_e$, [8] where $m_e$ is the free electron mass in vacuum. When **m**$_1$, **m**$_2$, **m**$_3$ are all within the *xz*-plane, we find that **T**$_3$ points along the *y*-direction. Figure S4 shows the magnitude $\tau_3$ of **T**$_3$ as the direction of **m**$_2$ rotates within the *xz*-plane while the direction of **m**$_3$ remains fixed, so that the angle $\theta_{23}$ changes. Note that $\tau_3$ is proportional to $\sin\theta_{23}$, in agreement with Eq. (S4). Importantly, we find that the amplitude of the $\sin\theta_{23}$ oscillation in $\tau_3$ depends on the direction of **m**$_1$, implying that there indeed exists a NNN interaction between F1 and F3. The three curves in Fig. S4 are obtained by simultaneously rotating **m**$_1$ and **m**$_2$, both within the *xz*-plane, while maintaining **m**$_1$·**m**$_2$=+1 (solid line), 0 (dash-dotted line), and -1 (dashed line), respectively. This dependence of $\tau_3$ on **m**$_1$·**m**$_2$ is in agreement with Eq. (S4). By comparing the three curves with Eq. (S4), we find $J_1M_s^2/A=14$ eV/μm$^2$ and $J_2M_s^2/A=3.5$ eV/μm$^2$, where $A$ is the interface area between F3 and N2. Thus the ratio $J_2/J_1$ becomes 0.24, which is very close to the value of 1/4 assumed in the main manuscript.

- **4-step transition process**

The 4-step transition occurs not only in the restoring process, but also in the saturation when the saturation is carried out after restoration without changing the field direction, as shown with open triangles in Fig. S5. The three stable resistance states of the second saturation, marked as SS1, SS2, and SS3, exactly match the states in the restoring process, indicating that the second saturation takes place via the same sequence of spin configurations that occurred in the restoring process (i.e., the AFM2→FM1 and FM1→AFM2 processes trace identical spin configurations in reverse order).

**Table S1**. Structural parameters of GaMnAs/GaAs:Be multilayers investigated in this work.

| Samples | dM | dN | Mn (%) | Number of periods |
|---|---|---|---|---|
| 1 | 8.0 / 28 | 4.0 / 14 | 5.5 | 8 |
| 2 | 8.0 / 28 | 4.0 / 14 | 3.5 | 7 |
| 3 | 8.0 / 28 | 4.0 / 14 | 3.5 | 8 |

(Parameters dM and dN denote thicknesses of GaMnAs and GaAs:Be layers, respectively, in units of nm/ML)

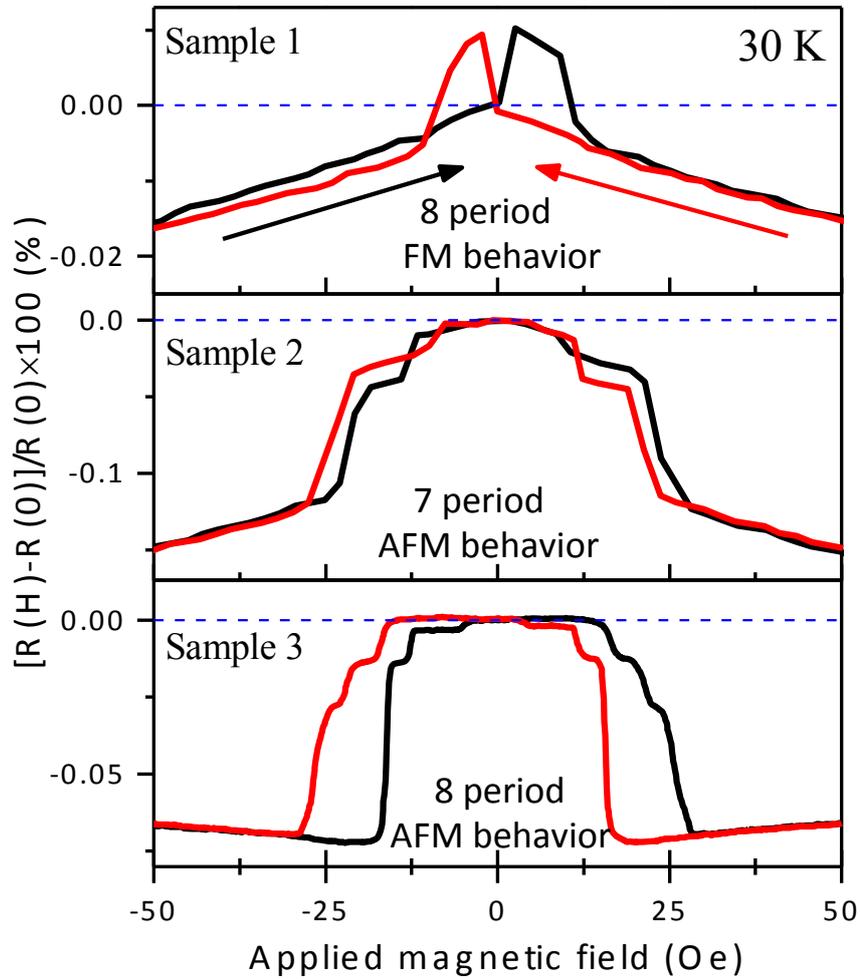

**Fig. S1.**
Magnetoresistance observed in three GaMnAs/GaAs multilayers with different structural parameters. Sample 1 shows only anisotropic magnetoresistance (AMR), characteristic of FM IEC between the GaMnAs layers; Samples 2 and 3 show a GMR-like behavior, indicating the presence of AFM IEC in these samples similar to that seen in the [GaMnAs/GaAs:Be]$_{10}$ multilayer discussed in detail in the main text.

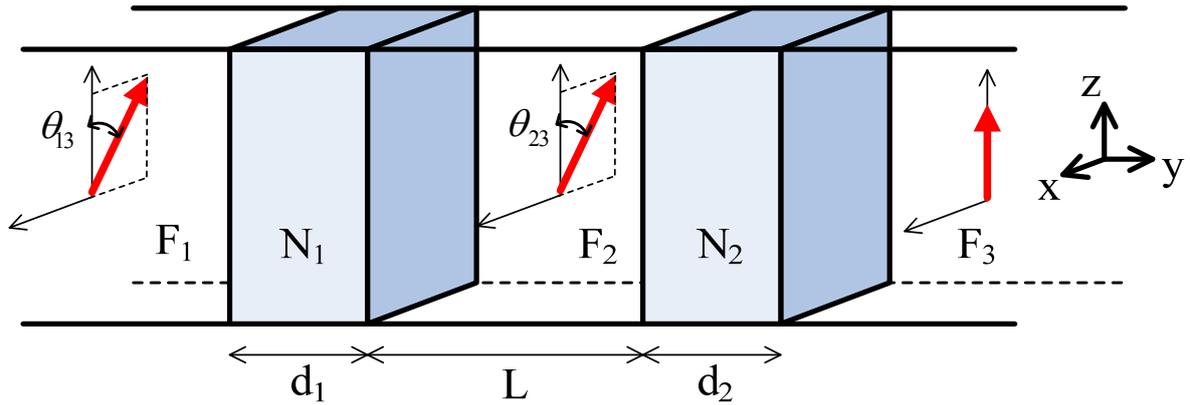

**Fig. S2**
Schematic figure of the model system. There are three magnetic layers with two non-magnetic spacers. The two magnetic layers, F1 and F3, are semi-infinite, and the other layers have finite thicknesses designated as $d_1$, $d_2$ and $L$. The direction of the magnetization of magnetic layer F3 is defined as the $z$-axis, and all magnetizations are assumed to be in the $xz$ plane.

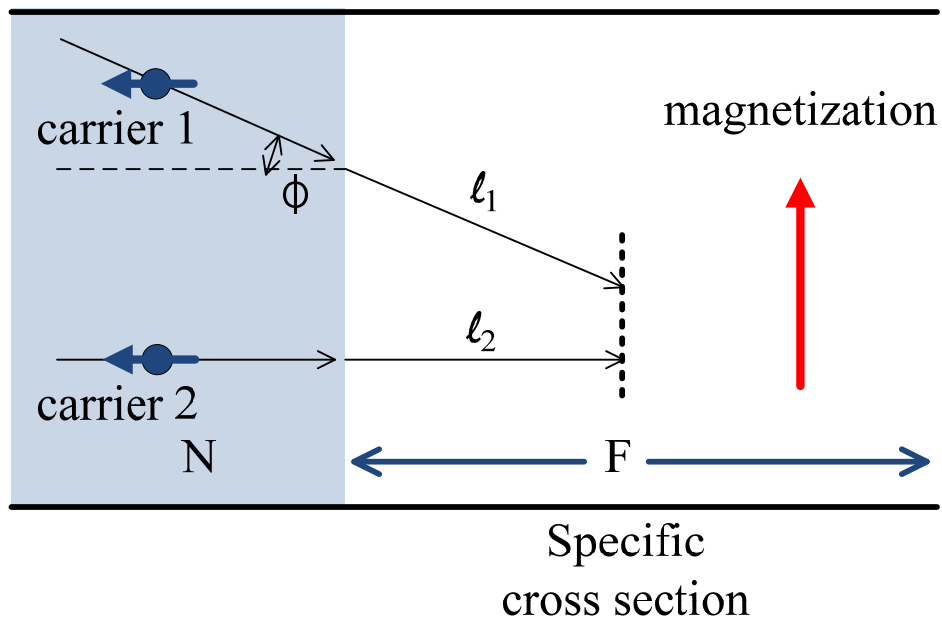

**Fig. S3**
Schematic representation of the spin dephasing process. Carrier 1 is incident from a nonmagnetic layer onto a magnetic layer with the incident angle $\phi$ and carrier 2 is normally incident. When the two carriers propagate into the magnetic layer by the same depth ($l_1 \cos\phi = l_2$), the distances traveled in the ferromagnetic layer are different for the two carriers: $l_1 > l_2$. Then, since the spin precession angle is proportional to the traveled distance, the spin precession angle during this travel is different for the two carriers, giving rise to spin dephasing.

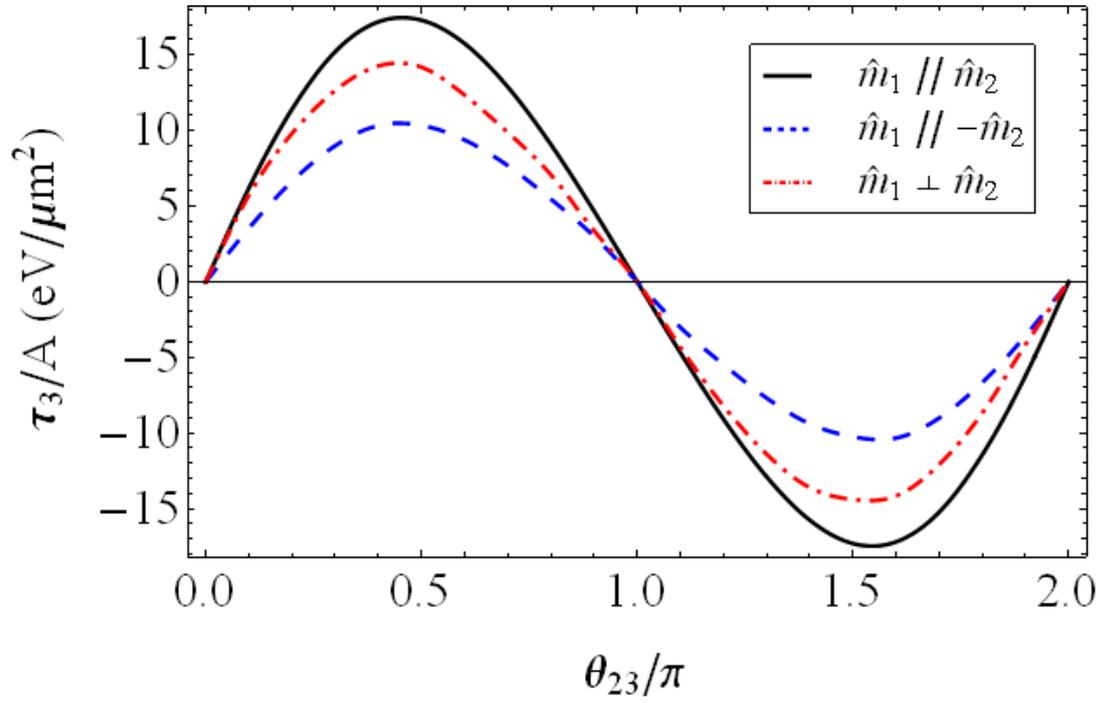

**Fig. S4**

The magnitude $\tau_3$ of the torque $\mathbf{T}_3$ on F3 as a function of $\theta_{23}$ (the angle between magnetizations in layers F2 and F3) for parallel ($\mathbf{m}_1//\mathbf{m}_2$), antiparallel ($\mathbf{m}_1//-\mathbf{m}_2$), and perpendicular ($\mathbf{m}_1 \perp \mathbf{m}_2$) configurations. All curves have sinusoidal angular dependences, but their amplitudes depend on $\mathbf{m}_1 \cdot \mathbf{m}_2$.

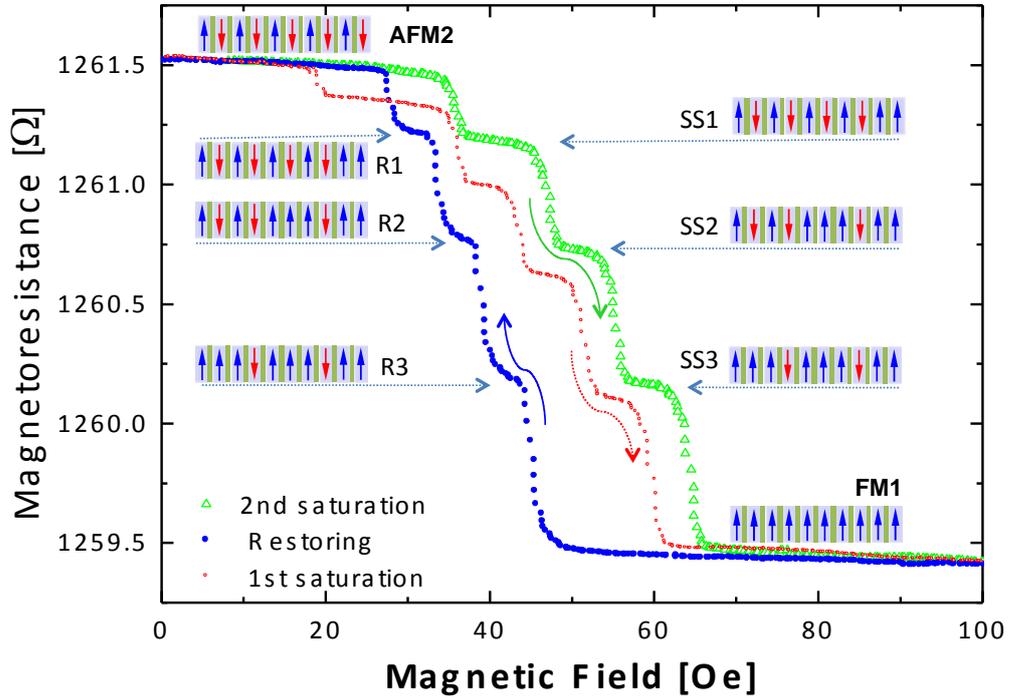

**Fig. S5**
Magnetoresistance hysteresis loop obtained by cycling the positive field from saturation to restoration and back again. The open circles (red), solid circles (blue), and open triangles (green) represent data from the initial saturation, restoration, and second saturation sweeps, respectively. The states realized in the second saturation process are marked as SS1-SS3 and the corresponding spin configurations are schematically shown in the figure. Unlike the initial 5-step saturation process, in which the first transition corresponds to the magnetization flip of the bottom GaMnAs layer, the second saturation is a *4-step process*, in which the first transition corresponds to the magnetization flip of the top GaMnAs layer. Both the restoration and the second saturation are 4-step processes, and exhibit identical spin configurations in reverse order.

**References and Notes**

[1] D. V. Baxter, D. Ruzmetov, J. Scherschligt, Y. Sasaki, X. Liu, J. K. Furdyna, and C. H. Mielke, Phys. Rev. B **65**, 212407 (2002).
[2] K. Hamaya, T. Taniyama, Y. Kitamoto, R. Moriya, and H. Munekata, J. Appl. Phys. **94**, 7657 (2003).
[3] S. S. P. Parkin, A. Mansour, and G. P. Felcher, Appl. Phys. Lett. **58**, 1473 (1991).
[4] Y. Yafet, Phys. Rev. B **36**, 3948 (1987).
[5] M. D. Stiles, and A. Zangwill, Phys. Rev. B **66**, 014407 (2002).
[6] D. C. Ralph, and M. D. Stiles, Journal of Magnetism and Magnetic Materials **320**, 1190 (2008).
[7] I. Theodonis, N. Kioussis, A. Kalitsov, M. Chshiev, and W. H. Butler, Phys. Rev. Lett. **97**, 237205 (2006).
[8] A. D. Giddings, T. Jungwirth, and B. L. Gallagher, Phys. Rev. B **78**, 165312 (2008).
[9] J. C. Slonczewski, Phys. Rev. B **39**, 6995 (1989).
[10] M. Dobrowolska K. Tivakornsasithorn, X. Liu, J. K. Furdyna, M. Berciu, K. M. Yu and W. Walukiewicz, Nature Mater. 11, 444 (2012).